\documentclass[aps,prd,showpacs,nofootinbib,superscriptaddress,preprintnumbers,
twocolumn
]{revtex4-1}
\usepackage{graphicx,amsmath,amssymb,soul,bbm}
\usepackage{dcolumn,booktabs,multirow,array,float}
\usepackage{enumitem,lmodern,dsfont,microtype,nicefrac}
\usepackage[dvipsnames]{xcolor}
\usepackage[utf8]{inputenc}
\usepackage{kantlipsum}
\usepackage{mathtools}
\usepackage{appendix}

\usepackage[compat=1.1.0]{tikz-feynman}
\usepackage[compat=1.1.0]{tikz-feynhand}
\usetikzlibrary{plotmarks}
\usetikzlibrary{shapes}
\usepackage{bm}
\usepackage[pdftex,
            pdftitle={Dalitz plots and lineshape of $a_1(1260)$},
            pdfauthor={Daniel Sadasivan, Maxim Mai, Hakan Akdag, Michael Döring},
            bookmarks,
            colorlinks,
            linkcolor=NavyBlue,
            citecolor=RubineRed,
            menucolor=black,
            urlcolor=NavyBlue,
            plainpages=false,
            pdfpagelabels,
            hypertexnames=false]{hyperref}

\newcommand{\cor}[2]{{\color{red}\st{#1}} {\color{blue} #2}}

\graphicspath{{plots/}}

\begin{document}

\title{
Dalitz plots and lineshape of \texorpdfstring{$a_1(1260)$}{} from a relativistic three-body unitary approach
}

\author{D.\ Sadasivan}
\email{sadasivan@email.gwu.edu}
\affiliation{The George Washington University, Washington, DC 20052, USA}

\author{M.\ Mai}
\email{maximmai@gwu.edu}
\affiliation{The George Washington University, Washington, DC 20052, USA}

\author{H.\ Akdag}
\email{akdag@hiskp.uni-bonn.de}
\affiliation{Helmholtz-Institut für Strahlen- und Kernphysik (Theorie), Universität Bonn, 53115 Bonn, Germany}

\author{M.\ D\"oring}
\email{doring@gwu.edu}
\affiliation{The George Washington University, Washington, DC 20052, USA}
\affiliation{Thomas Jefferson National Accelerator Facility, Newport News, VA 23606, USA}


\begin{abstract}
We formulate the final-state interaction of the $a_1(1260)$ resonance decay in a manifestly three-body unitary parameterization  and fit it to the $a_1(1260)$ lineshape measured by the ALEPH experiment. Dalitz plots calculated from this fit are presented. The work demonstrates the feasibility to numerically solve a previously derived amplitude and its generalization to isobars with spin and coupled channels. The model can also be applied to other meson decays and modified for the finite-volume problem as it arises in lattice QCD due to its manifest unitarity.
\end{abstract}

%
\maketitle

\section{Introduction}\label{sec:intro}

Excited states of strongly interacting matter exhibit a complex spectrum at low and intermediate energies. Connecting the underlying non-perturbative realization of Quantum Chromodynamics (QCD) to phenomenology is an ongoing experimental and theoretical challenge. The quest in understanding phenomenology including the decay of excited states but also unraveling the microscopic nature of mesonic and baryonic resonances poses open questions. Many are related to three-body dynamics. A prime example of such a riddle in the baryon sector (explored by CLAS@JLab~\cite{Ripani:2002ss, Aznauryan:2008pe}, ELSA~\cite{Sokhoyan:2015fra}, MAMI~\cite{Ahrens:2003na}, and other facilities) is the dynamical structure of the enigmatic Roper-state $N(1440)1/2^+$ and its mass pattern compared to the $N(1535)1/2^-$ state. While the microscopic features of the latter state can be parametrized by, e.g., meson-baryon interactions within Chiral Unitary approaches~\cite{Kaiser:1995cy,Inoue:2001ip,Nieves:2001wt, Doring:2009yv,Bruns:2010sv}, the situation is more involved for the Roper-state due to large branching ratios to the $\pi\pi N$ channels. The $\pi\pi N$ dynamics is included in many analyses of the baryon spectrum with varying degrees of rigor~\cite{Manley:1984jz, Arndt:2006bf, Mokeev:2008iw, Anisovich:2011fc}, with dynamical coupled-channel approaches usually respecting at least aspects of three-body unitarity~\cite{Kamano:2008gr, Doring:2009yv, Shklyar:2014kra}.
The unusually large branching ratios of the Roper resonance into three-body channels might explain not only its distorted shape~\cite{Arndt:2006bf} but also its unusual signals found in lattice QCD~\cite{Lang:2016hnn}. The present work deals with a simpler mesonic resonant three-body system, thus, providing a necessary stepping stone on the way to tackle more complex baryonic resonances in the future.

\setlength{\feynhandlinesize}{0.5mm}
\begin{figure*}[t]
\[
	\left(
	\begin{tikzpicture}[baseline=0cm]
	\begin{feynman}[baseline=(a0)]
	\begin{feynhand}
	\vertex 	(aa) {}; 
	\vertex [right=0.15cm of aa ](a0) {\(\vphantom{\Big|}\)};
	\vertex [above=0.065cm of aa] 	      (a0u){};
	\vertex [right=0.135cm of a0u] 	    (a0ur){};
	\vertex [below=0.065cm of aa] 	      (a0d){};
	\vertex [right=0.135cm of a0d] 	    (a0dr){};
	\vertex [right=1.55cm of a0,dot] 	   (c1){};
	\vertex [above=0.065cm of c1] 	     (c1u){};
	\vertex [below=0.065cm of c1] 	     (c1d){};
	\vertex [right=1.9cm of c1] 	    (d1){};
	\vertex [above=1cm of d1] 	    (d2){};
	\vertex [below=1cm of d1]       (d3){};
	
	\propag[very thick] (a0ur) to [edge label={$a_1(1260) \hphantom{.}$}] (c1u);
	\propag[very thick] (a0)  to (c1);
	\propag[very thick] (a0dr) to (c1d);
	\propag[double, very thick] (c1) to [out=45, in=180]   (d2);
	\propag[very thick]         (c1) to  [out=-45, in=180]  (d3);
	
	\filldraw[fill=Black, line width=0.2mm] (1.6,-0.125) rectangle (1.85,0.125);
	
	\node at (-0.06,0) [crossdot,  line width=0.4mm]{$\vphantom{\int\limits_{1}}$};
	\end{feynhand}
	\end{feynman}
	\end{tikzpicture}
	+
		\begin{tikzpicture}[baseline=0cm]
	\begin{feynman}[baseline=(a0)]
	\begin{feynhand}
	\vertex 	(aa) {}; 
	\vertex [right=0.15cm of aa ](a0) {\(\vphantom{\Big|}\)};
	\vertex [above=0.07cm of aa] 	      (a0u){};
	\vertex [right=0.135cm of a0u] 	    (a0ur){};
	\vertex [below=0.07cm of aa] 	      (a0d){};
	\vertex [right=0.135cm of a0d] 	    (a0dr){};
	\vertex [right=0.5cm of a0,dot] 	   (c1){};
	\vertex [above=0.07cm of c1] 	     (c1u){};
	\vertex [below=0.07cm of c1] 	     (c1d){};
	\vertex [right=1.9cm of c1] 	    (d1){};
	\vertex [above=1cm of d1] 	    (d2){};
	\vertex [below=1cm of d1]       (d3){};
	
	\propag[double, very thick] (c1) to [out=45, in=180]   (d2);
	\propag[very thick]         (c1) to  [out=-45, in=180]  (d3);
	
	\node at (0.7, 0) [grayblob, line width=0.4mm]  {};
	\end{feynhand}
	\end{feynman}
	\end{tikzpicture}
	\right)
	\left(
	\scalebox{1}{
		\begin{tikzpicture}[baseline=0cm]
		\begin{feynman}[baseline=(a0)]
		\begin{feynhand}
		\vertex []                   (a0){};
		\vertex [above=1cm of a0] 	(a0u){};
		\vertex [below=1cm of a0] 	(a0d){};
		\vertex [right=1.5cm of a0] 	   (x1){};
		\vertex [above=1cm of x1] 	   (x0u){};
		\vertex [below=1cm of x1] 	   (x0d){};
		\propag[double, very thick] (a0u) to  (x0u);
		\propag[very thick]         (a0d) to  (x0d);
		\filldraw[fill=Salmon, line width=0.4mm] (0.75,1) circle [radius=0.3cm];
		\node at (0.75,1)  {\large$\tau$};
		\end{feynhand}
		\end{feynman}
		\end{tikzpicture}
	}
	+
	\scalebox{1}{
		\begin{tikzpicture}[baseline=0cm]
		\begin{feynman}[baseline=(a0)]
		\begin{feynhand}
		\vertex []                   (a0){};
		\vertex [above=1cm of a0] 	(a0u){};
		\vertex [below=1cm of a0] 	(a0d){};
		\vertex [right=4.4cm of a0] 	   (x1){};
		\vertex [above=1cm of x1] 	   (x0u){};
		\vertex [below=1cm of x1] 	   (x0d){};
		\propag[double, very thick] (a0u) to  (x0u);
		\propag[very thick]         (a0d) to  (x0d);
		\filldraw[fill=SkyBlue, line width=0.7mm] (1.4,-1.2) rectangle (3,1.2)
		node[pos=0.5]  {\LARGE $T$};
		\filldraw[fill=Salmon, line width=0.4mm] (0.7,1) circle [radius=0.3cm];
		\node at (0.7,1)  {\large$\tau$};
		\filldraw[fill=Salmon, line width=0.4mm] (3.7,1) circle [radius=0.3cm];
		\node at (3.7,1)  {\large$\tau$};
		\end{feynhand}
		\end{feynman}
		\end{tikzpicture}
	}
	\right)
	\scalebox{1}{
		\begin{tikzpicture}[baseline=0cm]
		\begin{feynman}[baseline=(a0)]
		\begin{feynhand}
		\vertex [] (a0) {};
		\vertex [above=1cm of a0] 	(a0u){};
		\vertex [below=1cm of a0] 	(a0d){};
		\vertex [right=1.6cm of a0] 	(x0) {};
		\vertex [above=0.5cm of x0] (x0m) {\(\pi^+\)};
		\vertex [above=1.5cm of x0] (x0u) {\(\pi^-\)};
		\vertex [below=1cm of x0] 	(x0d) {\(\pi^-\)};
		\vertex [right=0.8cm of a0u,dot] 	(b){};
		\propag[very thick] (a0d) to (x0d);
		\propag[very thick] (b) to (x0m);
		\propag[very thick] (b) to (x0u);
		\propag[double, very thick] (a0u) to (b);
		\node at (b)  [above=0.1cm] {$v$};
		\end{feynhand}
		\end{feynman}
		\end{tikzpicture}
	}
\]
\caption{
\label{fig:decayprocess}
The decay of an unpolarized $a_1(1260)$ meson parametrized by a source (left parentheses) into a $\rho$-meson (double lines) and a pion (single lines), with subsequent decay into $\pi^-\pi^-\pi^+$ and the pertinent symmetrization (not shown). The relativistic, unitary final-state interaction (parentheses in the middle) is parametrized in terms of the isobar-spectator amplitude $T$ and isobar propagator $\tau$. The coupling of the $\rho$ to two pions is encoded in the vertex function $v$.} 
\end{figure*}

In the meson sector, the GlueX~\cite{AlGhoul:2017nbp} and COMPASS experiments~\cite{Alekseev:2009aa}, and the  BESIII accelerator~\cite{Asner:2008nq} search for the exited states of strongly interacting matter, including exotics, i.e., states with quantum numbers not possible to form from  two constituent quarks. Thus, it is expected that the discovery and the understanding of such states will be a direct indicator for gluonic degrees of freedom in QCD at low energies. Many of such exotic states, but also conventional mesons decay dominantly or exclusively into three particles, which has triggered extended partial-wave analysis (PWA) efforts, e.g., by COMPASS~\cite{Adolph:2015tqa, Akhunzyanov:2018lqa}, BESIII~\cite{Zou:2002ar, Ablikim:2004wn}, in coupled channels using the PAWIAN framework for $p\bar p$ induced meson production~\cite{Albrecht:2019ssa}, or using Khuri-Treiman equations and related frameworks by the Bonn group, JPAC, and others for light meson decays~\cite{Colangelo:2009db, Kubis:2009sb, Schneider:2010hs,
Kampf:2011wr, Niecknig:2012sj, Schneider:2012ez, Guo:2011aa,
MartinezTorres:2008kh, MartinezTorres:2011vh,
Danilkin:2014cra, Guo:2014vya, Guo:2015zqa,  Guo:2016wsi, Isken:2017dkw, Albaladejo:2017hhj, Hoferichter:2018kwz, Mikhasenko:2018bzm, Jackura:2018xnx, Gasser:2018qtg, Oset:2018zgc, Mathieu:2019fts,  Albaladejo:2019huw, Mikhasenko:2019rjf, Mikhasenko:2019vhk}. 
This paper extends the three-body approach of Ref.~\cite{Mai:2017vot} that is inspired by work of Amado et al.~\cite{Aaron:1969my} (see Sec.~\ref{sec:formalism} for a detailed discussion).

Lattice QCD is the non-perturbative tool to access the QCD Greens functions at low and intermediate energies from first principles. In addressing three-hadron systems, such calculations need to include a large set of operators including three-hadron operators. This leads to a significantly increased computational effort compared to two-body systems. However, progress has been made over the last years in calculating lattice spectra~\cite{Beane:2007es,Lang:2016hnn,Lang:2014tia, Horz:2019rrn, Woss:2019hse, Culver:2019vvu}, including the $a_1(1260)$ meson~\cite{Lang:2014tia}. Such discrete and real-valued spectra are inherently different from the infinite volume ones. The so-called quantization condition allows one to map between finite and infinite volume, and is an active field of theoretical research~\cite{Sharpe:2017jej,Hammer:2017kms, Hammer:2017uqm,Guo:2017ism, Guo:2016fgl,Hansen:2016ync,Hansen:2016fzj, Hansen:2015zta,Hansen:2015zga,Meissner:2014dea,Hansen:2014eka,Briceno:2012rv,Kreuzer:2012sr, Roca:2012rx, Polejaeva:2012ut,Kreuzer:2008bi,Mai:2017bge,Doring:2018xxx,Mai:2018xwa,Hansen:2019nir}. 
It has been shown in Ref.~\cite{Mai:2017bge} that the key to the understanding of three-body finite-volume spectra from the lattice lies in the $S$-matrix principle of unitarity. There, a simpler version of the framework underlying the present study~\cite{Mai:2017vot} (no spin) was adopted to the finite volume demonstrating its feasibility to provide infinite-volume mappings. 
Both finite-volume and lattice computations in the three-body sector are now capable of addressing simple three-pion systems~\cite{Culver:2019vvu,Blanton:2019vdk,Mai:2018djl,Mai:2019fba} and are about to be extended to more complicated cases like axial mesons and exotics. 

The $a_1(1260)$ axial meson has a clean three-pion decay as the intermediate state in $\tau$ decays in which other partial waves are suppressed, in contrast to pion or photon-induced three-pion production; it is also wide~\cite{Tanabashi:2018oca} indicating strong and non-trivial three-body effects which makes the $a_1(1260)$ a prime candidate to study three-body dynamics. This is reflected in an increased interest in the structure of the $a_1(1260)$~\cite{Janssen:1993nj, Lutz:2003fm, Roca:2005nm, Wagner:2007wy, Wagner:2008gz, Lutz:2008km, Kamano:2011ih, Nagahiro:2011jn, Zhang:2018tko, Dai:2018zki, Mikhasenko:2018bzm} which is also the main goal of the present manuscript.

More specifically, in view of the importance of three-body unitarity, our goal is to extend the manifestly unitary, relativistic three-body scattering amplitude derived in Refs.~\cite{Mai:2017vot,Mai:2017wdv} to the coupled-channel case for isobars with spin, namely the $a_1(1260)$ in which the $\rho\pi$ decay channel is known to provide the dominant contribution~\cite{Tanabashi:2018oca}. This also means the development of numerical techniques for the solution of the integral equations. Here, we restrict the $a_1(1260)$ dynamics to the $\rho\pi$ $S$- and $D$-wave channels; a detailed partial-wave analysis of the three-pion system in $\tau$ decays measured by CLEO~\cite{Asner:1999kj} has shown that there are also other channels needed for the detailed description of Dalitz plots and related observables. Demonstrating the feasibility of the approach, we fit the amplitude to the experimental data on the $\tau^- \rightarrow \pi^+\pi^-\pi^- \nu_\tau$ lineshape measured in the ALEPH$@$CERN~\cite{Schael:2005am} experiment. Note also other measurements of the same process~\cite{Albrecht:1992ka,Akers:1995vy,Abreu:1998cn} with lower statistics.

This work is organized as follows: In Sec.~\ref{sec:formalism} the main definitions of the decay and three-body amplitude are introduced and compared to other approaches in the literature. The strategy for solving these equations is described in Sec.~\ref{subsec:implementation}. Finally, in Sec.~\ref{subsec:numresults} the result of a fit to ALEPH data and the calculation of Dalitz plots from that fit will be presented and discussed.

\setlength{\feynhandlinesize}{0.5mm}
\begin{figure*}[t]
\[
\scalebox{0.8}
{
\begin{tikzpicture}[baseline=0cm]
  \begin{feynman}[baseline=(a0d)]
  \begin{feynhand}
    \vertex [] 	                      (a0){};
    \vertex [below=1cm of a0] 	      (ad){};
    \vertex [above=1cm of a0] 	      (au){};

    \vertex [right=2cm of a0] 	    (b0){};
    \vertex [below=1cm of b0] 	      (bd){};
    \vertex [above=1cm of b0] 	      (bu){};

    \vertex [right=2cm of b0]       (x0){};
    \vertex [below=1cm of x0] 	      (xd){};
    \vertex [above=1cm of x0] 	      (xu){};
    \propag[double, very thick] (au) to (bu);
    \propag[double, very thick] (bd) to (xd);
    \propag[very thick]         (ad) to (bd);
    \propag[very thick]         (bu) to (xu);
    \filldraw[fill=SkyBlue, line width=0.7mm] (1.2,-1.2) rectangle (2.8,1.2)
    node[pos=0.5]  {\LARGE $T$};
 \end{feynhand}
 \end{feynman}
\end{tikzpicture}
}
=
\left(
\scalebox{0.8}
{
\begin{tikzpicture}[baseline=0cm]
  \begin{feynman}[baseline=(ad)]
  \begin{feynhand}
    \vertex [] 	                      (a0){};
    \vertex [below=1cm of a0] 	      (ad){};
    \vertex [above=1cm of a0] 	      (au){};

    \vertex [right=0.7cm of a0] 	      (b0){};
    \vertex [below=1cm of b0] 	      (bd){};
    \vertex [above=1cm of b0,dot] 	      (bu){};

    \vertex [right=0.7cm of b0] 	      (c0){};
    \vertex [below=1cm of c0,dot] 	      (cd){};
    \vertex [above=1cm of c0] 	      (cu){};

    \vertex [right=0.7cm of c0]         (x0){};
    \vertex [below=1cm of x0] 	      (xd){};
    \vertex [above=1cm of x0] 	      (xu){};
    \propag[double, very thick] (au) to (bu);
    \propag[double, very thick] (cd) to (xd);
    \propag[very thick]         (bu) to (cd);
    \propag[very thick]         (ad) to (cd);
    \propag[very thick]         (bu) to (xu);
 \end{feynhand}
 \end{feynman}
\end{tikzpicture}
}
+
\scalebox{0.8}
{
\begin{tikzpicture}[baseline=0cm]
  \begin{feynman}[baseline=(a0)]
  \begin{feynhand}
    \vertex [] 	                      (a0){};
    \vertex [below=1cm of a0] 	      (ad){};
    \vertex [above=1cm of a0] 	      (au){};

    \vertex [right=2cm of a0]         (b0){};
    \vertex [below=1cm of b0] 	      (bd){};
    \vertex [above=1cm of b0] 	      (bu){};
    \propag[double, very thick] (au) to (bd);
    \propag[very thick]         (ad) to (bu);

  \node[draw,scale=0.9,diamond,fill=orange!50] at (1,0)  {\bf$C$};
 \end{feynhand}
 \end{feynman}
\end{tikzpicture}
}
\right)
+
\left(
\scalebox{0.8}
{
\begin{tikzpicture}[baseline=0cm]
  \begin{feynman}[baseline=(ad)]
  \begin{feynhand}
    \vertex [] 	                      (a0){};
    \vertex [below=1cm of a0] 	      (ad){};
    \vertex [above=1cm of a0] 	      (au){};

    \vertex [right=0.7cm of a0] 	      (b0){};
    \vertex [below=1cm of b0] 	      (bd){};
    \vertex [above=1cm of b0,dot] 	      (bu){};

    \vertex [right=0.7cm of b0] 	      (c0){};
    \vertex [below=1cm of c0,dot] 	      (cd){};
    \vertex [above=1cm of c0] 	      (cu){};

    \vertex [right=0.7cm of c0]         (x0){};
    \vertex [below=1cm of x0] 	      (xd){};
    \vertex [above=1cm of x0] 	      (xu){};
    \propag[double, very thick] (au) to (bu);
    \propag[double, very thick] (cd) to (xd);
    \propag[very thick]         (bu) to (cd);
    \propag[very thick]         (ad) to (cd);
    \propag[very thick]         (bu) to (xu);
 \end{feynhand}
 \end{feynman}
\end{tikzpicture}
}
+
\scalebox{0.8}
{
\begin{tikzpicture}[baseline=0cm]
  \begin{feynman}[baseline=(a0)]
  \begin{feynhand}
    \vertex [] 	                      (a0){};
    \vertex [below=1cm of a0] 	      (ad){};
    \vertex [above=1cm of a0] 	      (au){};

    \vertex [right=2cm of a0]         (b0){};
    \vertex [below=1cm of b0] 	      (bd){};
    \vertex [above=1cm of b0] 	      (bu){};
    \propag[double, very thick] (au) to (bd);
    \propag[very thick]         (ad) to (bu);

  \node[draw,scale=0.9,diamond,fill=orange!50] at (1,0)  {\bf$C$};
 \end{feynhand}
 \end{feynman}
\end{tikzpicture}
}
\right)
\scalebox{0.8}
{
\begin{tikzpicture}[baseline=0cm]
  \begin{feynman}[baseline=(a0)]
  \begin{feynhand}
    \vertex [] 	                      (a0){};
    \vertex [below=1cm of a0] 	      (ad){};
    \vertex [above=1cm of a0] 	      (au){};

    \vertex [right=2cm of a0] 	    (b0){};
    \vertex [below=1cm of b0] 	      (bd){};
    \vertex [above=1cm of b0] 	      (bu){};

    \vertex [right=2cm of b0]       (x0){};
    \vertex [below=1cm of x0] 	      (xd){};
    \vertex [above=1cm of x0] 	 (xu){};
    \propag[double, very thick] (ad) to (bd);
    \propag[double, very thick] (bu) to (xu);
    \propag[very thick]         (au) to (bu);
    \propag[very thick]         (bd) to (xd);
    \filldraw[fill=SkyBlue, line width=0.7mm] (1.2,-1.2) rectangle (2.8,1.2)
    node[pos=0.5]  {\LARGE $T$};
    \filldraw[fill=Salmon, line width=0.4mm] (0.7,-1) circle [radius=0.3cm];
    \node at (0.7,-1)  {\large$\tau$};
 \end{feynhand}
 \end{feynman}
\end{tikzpicture}
}
\]
\caption{
\label{fig:T-matrix}
The isobar-spectator amplitude leading to a unitary three-pion scattering amplitude. The interaction kernel (quantity in parentheses) is determined from unitarity and comprises a complex valued contribution shown as one-pion exchange, as well as a real valued three-body force $C$.}
\end{figure*}

\section{Formalism}\label{sec:formalism}

The final-state interaction of the weakly induced decay process ${\tau^- \rightarrow \pi^+\pi^-\pi^- \nu_\tau}$ is given by the interaction of three pions with $a_1(1260)$ quantum numbers. The presence of the outgoing neutrino allows the total energy squared, $s$, of the three pions to vary. This allows one to ``scan'' the spectrum of three pions in the final state, thus obtaining the so-called \textit{mass spectrum} or \textit{lineshape} of the $a_1(1260)$-resonance that was measured in the ALEPH~\cite{Schael:2005am} experiment and is the main experimental input of the present work.

To access the mass spectrum theoretically, the decay process is decomposed into the weak and strong parts as ${\tau^-\rightarrow W^-\nu_\tau\rightarrow 
(a_1(1260)\rightarrow\pi^-\pi^-\pi^+)\nu_\tau}$ (see, e.g., Ref.~\cite{Mikhasenko:2018bzm}). The strong, final-state interaction of the three pions is described by the process shown in Fig.~\ref{fig:decayprocess}. As indicated in the first parentheses it contains a part describing the production of the $\rho\pi$ pair in $S$- and $D$-wave. This part consists of the direct production of the $\rho\pi$ pair as well as an intermediate propagation of $a_1(1260)$. The $\rho\pi$ system interacts then (second parentheses in the figure) and decays in the final step into three pions, such that three-body unitarity is preserved exactly. Note that while the picture suggests a diagrammatic expansion of the interaction, the approach is not Lagrangian-based as discussed below.

The final-state interaction of three pions is taken into account non-perturbatively ensuring three-body unitarity. This is achieved using the formalism developed in Ref.~\cite{Mai:2017vot}.  In a nutshell, the approach is based on a decomposition of the scattering amplitude into a connected and a disconnected part (cf. "connectedness structure" in Ref.~\cite{Eden:1966dnq}). Each of these pieces is populated by the two-body sub-system (referred to as \textit{``isobar''}  in the following) and a spectator. The analytic forms of the Bethe-Salpeter kernel, $B$, and  the fully dressed isobar propagator $\tau$ are fixed up to real functions of energy and momenta by matching the Bethe-Salpeter equation with the three-body unitarity condition~\cite{Mai:2017vot}. This isobar-spectator amplitude is depicted symbolically in Fig.~\ref{fig:T-matrix}.

Note that this approach relies on dispersive techniques making advantage of unitarity and connectedness structure. As such, it does not rely on a Lagrangian formalism for the microscopic interaction but provides a clean separation between on-shell parts and short-range physics encoded in real-valued contributions indicated as $C$ in Fig.~\ref{fig:T-matrix} and referred to as ``\textit{three-body force}'' in what follows. 
We will describe the implementation and numerical applications of this approach in more detail below.

The formal decomposition of the entire amplitude into a short-range and a long-range part (``ladders'') is discussed in Ref.~\cite{Mikhasenko:2019vhk}. It is sometimes referred to as ``two-potential formalism'' used mostly in the baryonic sector~\cite{Matsuyama:2006rp, Ronchen:2012eg}. This decomposition is not unique~\cite{Doring:2009bi} but can be advantageous for time-consuming fits to large data sets~\cite{Ronchen:2012eg}.
The matching of different three-body formalisms, including the current one and its mapping to Feynman-diagrammatic expressions is discussed in Ref.~\cite{Jackura:2019bmu}. In that reference, global analytic properties of the three-to-three amplitude were discussed. Comparing frameworks like the present one to the analytic properties of the triangle diagram it was shown that in the sub-threshold region, non-analyticities can occur depending, e.g., on specific implementations of the integration over two-body sub-energies in the three-body equation.

Early attempts to solve the three-pion problem in the $a_1(1260)$ channel with a non-perturbative final-state interaction, in a framework similar to the present one, were carried out in Refs.~\cite{Janssen:1993nj, Janssen:1994uf}, sharing technical details like complex-momentum integration with our approach, but modeling the short-range $\pi\rho$ interaction with effective Lagrangians. We do not attempt to microscopically resolve this dynamics which can only be done model-dependently, anyways. The amplitude of Refs.~\cite{Janssen:1993nj, Janssen:1994uf} is initially formulated including an unstable $\rho$ isobar, as in the present case. However, due to problems of how to continue the solution of the $T$-matrix from complex back to real spectator momenta, the actual numerical results were obtained with a stable $\rho$ meson  which violates unitarity.
Indeed, the the isobar representing the two-body sub-amplitude must possess its proper imaginary part for the entire amplitude to fulfill three-body unitarity~\cite{Mai:2017vot}. This cannot be achieved with a stable particle propagator.

The present approach is most closely related to the one by the EBAC collaboration (now ANL/Osaka)~\cite{Kamano:2011ih}. There, Dalitz plots not only for the $a_1(1260)$ meson but also for other three-body  decays were predicted. Furthermore, inelasticities in the two-body subsystems were taken into account and the $a_1(1260)$ was allowed to decay also in $\sigma\pi$ and $f_2\pi$, apart from the dominant $\rho\pi$ $S$-wave channel. On the other hand, we use here an amplitude that is manifestly unitary, with the full proof of unitarity first delivered in Ref.~\cite{Mai:2017vot}, and we also provide data fits and study the energy dependence of the amplitude  by comparing to ALEPH data. This is particularly relevant for future extensions to finite-volume calculations for lattice QCD in which manifest unitarity is responsible for a subtle cancellation of unphysical singularities~\cite{Mai:2017bge}.

In the following we describe our approach to calculate the decay process. In Sec.~\ref{subsec:PlaneWave}, we provide the equations needed to describe each term shown in Fig.~\ref{fig:decayprocess}. In Sec.~\ref{subsec:PWA} the partial wave decomposition is discussed. Finally, in Sec.~\ref{subsec:Observables}, we describe how the amplitude can be related to observables.

\subsection{Plane-Wave Amplitudes}\label{subsec:PlaneWave}

The amplitude $\hat\Gamma_{\Lambda\lambda}$, describing the decay of the axial $a_1(1260)$-resonance at rest with helicity $\Lambda$ measured along the $z$-axis into a $\pi^-$ and a $\rho_\lambda^0\to \pi^+\pi^-$ with helicity $\lambda$, is given by
\begin{align}
\label{eq:decayrate}
&\hat \Gamma_{\Lambda\lambda} (\boldsymbol{q}_1,\boldsymbol{q}_2,\boldsymbol{q}_3)
=\frac{1}{\sqrt{2}}\big[\Gamma_{\Lambda\lambda} (\boldsymbol{q}_1,\boldsymbol{q}_2,\boldsymbol{q}_3)
-\left(\boldsymbol{q}_1\leftrightarrow\boldsymbol{q}_2\right)\big],
\nonumber \\
&\Gamma_{\Lambda\lambda} (\boldsymbol{q}_1,\boldsymbol{q}_2,\boldsymbol{q}_3)
= 
\\
&
\Big(D_{\Lambda\lambda}(\boldsymbol{q}_1)+
\sum_{\lambda'}\int \frac{d^3\boldsymbol{p}}{(2\pi)^3 2E_{p}} \,
D_{\Lambda\lambda'}(\boldsymbol{p})
\tau(\sigma(|\boldsymbol{p}|))
T_{\lambda'\lambda}(\boldsymbol{p},\boldsymbol{q}_1)
\Big)
\nonumber\\
&\qquad\qquad\qquad\qquad\qquad\qquad\qquad
\times\tau(\sigma(|\boldsymbol{q}_1|))v^{\pm}_{\lambda}(q_2,q_3)\,,
\nonumber
\end{align}
where $\bm{q}_1$, and $\bm{q}_2$ are outgoing $\pi^-$ momenta, and $\bm{q}_3$ is the outgoing $\pi^+$ momentum. In Eq.~(\ref{eq:decayrate}) the dependence on the squared invariant mass of the three-body system $s$ is suppressed; isospin indices are provided below;
the squared invariant mass of the $\rho$, denoted by $\sigma(|\bm{p}|)$, depends only on the size of the spectator momentum $\bm{p}$ and is abbreviated as $\sigma(p)\equiv \sigma(|\bm{p}|)$ in the following; confusion with four-vector notation should be excluded from the context.
Furthermore, $\sigma(p)=(P_3-p)^2=s + m_\pi^2 - 2 \sqrt{s}\,E_p$ with $E_p^2=m_\pi^2+\bm{p}^2$ and $P_3=(\sqrt{s},\bm 0)$ being the total four-momentum of three pions. Note that throughout this paper we are working in the three-body center-of-mass frame unless stated otherwise. Due to Bose symmetry, the amplitude is symmetric under exchange of the two $\pi^-$. The process is of odd intrinsic parity and the isospin part of the wave function contains an additional minus sign under the exchange of the two $\pi^-$, restoring the overall Bose symmetry. Some of the terms appearing in Eq.~\eqref{eq:decayrate} are shown in Fig.~\ref{fig:decayprocess} and are defined in the following. Note also that this three-dimensional relativistic equation with all pions on their respective mass shells emerges after carrying out the integration of zeroth dimension, applying the delta-distribution of the spectator from the isobar-spectator propagation, see  Ref.~\cite{Mai:2017vot}.

The elementary process $a_1 \rightarrow \pi \rho$ is indicated as $D_{\Lambda\lambda}$. It can have scalar or derivative character,
\begin{align}
D_{\Lambda\lambda}(\boldsymbol{q}_1)=
\epsilon_{\Lambda,\mu}\epsilon^{*\mu}_\lambda(\boldsymbol{q}_1)(-m_{a_1}g_{s}I_{a_1\rho\pi}+\dots)\,,
\label{eq:decaytower}
\end{align}
i.e., the $a_1\rho\pi$ coupling is given by a tower of Lagrangians which we do not aim to explicitly use in this study, including masses ($m_{a_1}$)  coupling constants ($g_s$) and isospin factors ($I_{a_{1}\rho\pi}$). Instead, we know that symmetry allows for even $\rho\pi$ partial waves each with their own energy and momentum dependence, which we simply parametrize in the angular momentum basis, further abbreviated with $JLS$ basis,  directly, where $JLS$ stand for total, orbital, and spin angular momenta, respectively. See Sec.~\ref{subsec:PWA} for a detailed discussion.

The vertex $v_{\lambda}^{\pm}$ in Eq.~(\ref{eq:decayrate}) is an elementary part of the $\rho^0$-decay  into a $\pi^+\pi^-$ pair with four-momenta $q_2$ and $q_3$,
\begin{align}
\label{eq:finalvertex}
v^{\pm}_{\lambda}(q_2,q_3)&=I'v_{\lambda}(q_2,q_3)\\
v_{\lambda}(q_2,q_3)&=
-ig_1\epsilon_{\lambda}^\mu(\boldsymbol{q}_1)~(q_2-q_3)_\mu F((q_2+q_3)^2,(q_2-q_3)^2)
\,,\nonumber
\end{align}
where $g_1$  is the $\rho\pi\pi$ coupling, $v_\lambda$ is the isospin-projected decay vertex, and $I'$ describes the transition from isospin to particle basis as needed only in the final $\rho$ decay. Note that the latter factor is irrelevant as long as there is only one isobar ($\rho^0$). Then this factor can be reabsorbed into the overall normalization of the $a_1$ decay. Furthermore, due to the azimuthal symmetry of the isobar momentum, the imaginary part of $\epsilon_\lambda$ does not contribute to the partial-wave projected $v_\lambda(q_2,q_3)$. Thus, in this case, the vertex is strictly real, i.e., $v_\lambda(q_2,q_3)=~v^*_\lambda(q_2,q_3)$. Any additional momentum dependence of the isobar decay is encoded in the covariant form-factor $F$ that is introduced to regularize ultraviolet divergences in the two- and three-body sector. Note that three-body unitarity imposes a consistent use of the form-factor in the self-energy $\tau$ and exchange $B$, which requires it to be covariant~\cite{Mai:2017vot}.  The choice of this real-valued functions is not unique. Explicit expressions used in this work are discussed in  Appendix~\ref{app:formfac}.

The three-body unitarity underlying this formalism implements two-body unitarity automatically~\cite{Mai:2017vot}. This constrains the imaginary part of the inverse of the isobar-spectator propagator. The remaining freedom allows one to pick a form of the isobar propagator $\tau$ that is suitable for the problem at hand. For example, a two-body scattering amplitude motivated by Chiral Perturbation Theory was implemented into the three-body scattering equation in Refs.~\cite{Mai:2018xwa,Mai:2018djl}. This form is very useful for repulsive or weakly attractive channels (such as, e.g., $\pi\pi$ scattering in the channel of the atypical $f_0(500)$-resonance). For the present work, dealing exclusively with the $\rho$-resonance, it is, however, justified to simply choose
\begin{align}
\tau(\sigma(p))=
\frac{1}{\sigma(p)-m_{\rho}^2-\Sigma(\sigma(p))}\,
\label{eq:taumatrix}
\end{align}
as represented in Fig.~\ref{fig:TauMatrix}.
This particular parametrization of the vector-isovector channel provides a sufficiently accurate representation of the physical on-shell two-body amplitude. For alternative forms of $\tau$, e.g., with two subtractions, see Ref.~\cite{Mai:2017vot}.
Here, $m_\rho$ is a real-valued free parameter not fixed by unitarity. We fit it, together with the coupling $g_1$ from Eq.~\eqref{eq:finalvertex} and the form-factor to two-body $\pi\pi$ phase-shift data~\cite{Protopopescu:1973sh,Estabrooks:1974vu}. Note that the $\tau$ in Eq.~(\ref{eq:taumatrix}) is the same as $S$ in Ref.~\cite{Mai:2017vot}. In Eq.~\eqref{eq:taumatrix}, $\Sigma$ is referred to as the self-energy. We use the explicit expression
\begin{align}
\Sigma(\sigma(p))=\int\limits_0^\infty \frac{dkk^2}{(2\pi)^3 E_k} \frac{ \tilde v(\sigma(p),k)\tilde v^*(\sigma(p),k) }{ \sigma(p)-4E_k^2+i\epsilon} \,.
\label{eq:selfenergy}
\end{align}
This can be evaluated in the isobar rest frame with the two-body 4-momentum  $P_2=(\sqrt{\sigma(p)},{\bf 0})$. The tilde on $\tilde v$ indicates that this vertex is projected to the total angular momentum of the $\rho$-meson as
\begin{align}
\tilde v(\sigma(p),k)= \sqrt{\frac{16\pi}{3}}g_1 k \tilde F(\sigma(p),k) \,,
\label{eq:projected-vertex}
\end{align}
which is derived from the $\rho\pi\pi$ vertex in Eq.~\eqref{eq:finalvertex}. The form-factor $\tilde F$ used here is of the same analytic structure as the $F$ in Eq.~\eqref{eq:finalvertex} for the on-shell region. It is, however, modified for technical reasons in the off-shell region, which does not violate unitarity. See Eq.~\eqref{eq:form2} in Appendix~\ref{app:formfac} for further details.
%

\setlength{\feynhandlinesize}{0.5mm}
\begin{figure}[t]
\centering
\[
\scalebox{1}{
\begin{tikzpicture}[baseline=0cm]
  \begin{feynman}[baseline=(a0)]
  \begin{feynhand}
    \vertex []                   (a0){};
    \vertex [above=0.001cm of a0] 	(a0u){};
  
   \vertex [right=2cm of a0] 	   (x1){};
   \vertex [above=0.001cm of x1] 	   (x0u){};
 
    \propag[double, very thick] (a0u) to  (x0u);
   
    \filldraw[fill=Salmon, line width=0.4mm] (1,0) circle [radius=0.3cm];
    \node at (1,0)  {\large$\tau$};
 \end{feynhand}
 \end{feynman}
\end{tikzpicture}
}
=
\scalebox{1}{
\begin{tikzpicture}[baseline=0cm]
  \begin{feynman}[baseline=(a0)]
  \begin{feynhand}
    \vertex []                    (a0){};
    \vertex [above=0.001cm of a0] (a0u){};
    \vertex [right=1.7cm of a0] 	  (x1){};
    \vertex [above=0.001cm of x1] (x0u){};
    \propag [double, very thick]  (a0u) to  (x0u);
 \end{feynhand}
 \end{feynman}
\end{tikzpicture}
}
+
\scalebox{1}{
\begin{tikzpicture}[baseline=0cm]
  \begin{feynman}[baseline=(a0)]
  \begin{feynhand}
    \vertex []                   (a0){};
    \vertex [above=0.001cm of a0] 	(a0u){};
   
    \vertex [right=0.7cm of a0] 	   (b1){};
    \vertex [above=0.001cm of b1,dot] 	   (b0u){};
    \vertex [right=1.3cm of a0] 	   (c1){};
    \vertex [above=0.001cm of c1,dot] 	   (c0u){};
    \vertex [right=3.3cm of a0] 	   (x1){};
    \vertex [above=0.001cm of x1] 	   (x0u){};
   
    \propag[double, very thick] (a0u) to  (b0u);
    \propag[double, very thick] (c0u) to  (x0u);

    \filldraw[fill=none, line width=0.4mm] (1,0) circle [radius=0.3cm];
    \node at (1,0);
    
    \filldraw[fill=Salmon, line width=0.4mm] (2.3,0) circle [radius=0.3cm];
    \node at (2.3,0)  {\large$\tau$};
    
	\node at (1,0)  [above=0.3cm] {$\Sigma$};
 \end{feynhand}
 \end{feynman}
\end{tikzpicture}
}
\]
\caption{
    \label{fig:TauMatrix}
    A diagrammatic depiction of the isobar propagator 
    from Eqs.~\eqref{eq:taumatrix} and \eqref{eq:selfenergy}, representing the re-summation of $\pi\pi$ interactions within the isobar to all orders. Double and single lines denote the isobar and stable particle propagators, respectively.
    }
\end{figure}

The plane wave solution for the isobar-spectator amplitude is obtained as described in Ref.~\cite{Mai:2017vot}, i.e., by solving the  Bethe-Salpeter type relativistic integral equation
\begin{align}
\label{eq:Bethe-Salpeter}
&T_{\lambda'\lambda}({\bm p},{\bm q}_1)=\left(B_{\lambda'\lambda}({\bm p},{\bm q}_1)+C\right)+
\\
&\sum_{\lambda''}\int  \frac{d^3\bm l}{(2\pi)^3 2E_l} \left(B_{\lambda'\lambda''}({\bm p},{\bm  l })+C\right)\tau(\sigma(l)) T_{\lambda''\lambda}({\bm l},{\bf q_1})\nonumber 
\,,
\end{align}
where $\lambda'$, $\lambda$ and $\lambda''$ are the helicities of the in-, outgoing, and intermediate $\rho$-isobar, and ${\bm p}$, ${\bm q}_1$, and $\bm l$ are the in-, outgoing and intermediate spectator momenta, respectively. The diagrammatic representation of this equation is shown in Fig.\ref{fig:T-matrix}. The first part of the driving term of Eq.~(\ref{eq:Bethe-Salpeter}), the so-called $B$-term, can be interpreted as the one-pion exchange process depicted in the first term inside the parentheses in Fig.~\ref{fig:T-matrix}. In the helicity and isospin bases it reads
\begin{align}
\label{eq:BmatrixPlane}
B_{\lambda'\lambda}(&{\bm p},{\bm q}_1)=\\
&\frac{(-1)\, v_{\lambda'}(P_3-p-q_1,q_1)v^*_{\lambda}(p,P_3-p-q_1)}{(\sqrt{s}-E_p-E_{q_1})^2-(m_\pi^2+p^2+q_1^2+2pq_1z)+i\epsilon} 
\ ,\nonumber
\end{align}
with an additional minus sign from the overall isospin factor of this process.
The  equation emerges from the sum of a forward and backward (in time) pion exchange between two $\rho\pi\pi$ vertices.  We denote the angle between the in- and outgoing isobars by $\theta_1$ and $z=\cos{\theta_1}$. The second part of the driving term of the integral equation~\eqref{eq:Bethe-Salpeter} is a real-valued function $C$ of spectator momenta and $s$, i.e., $C\equiv C(\boldsymbol{p},\boldsymbol{q},s) $. It arises from the fact that only imaginary parts of the amplitude are fixed as discussed in Ref.~\cite{Mai:2017vot}. Physically, this function is related to the three-body contact term via a decay of the isobar into two in- and outgoing pions, see e.g. Ref.~\cite{Jackura:2019bmu}. To capture the $\rho\pi$ contact interaction efficiently, we will model this function in the next section after projecting the whole integral equation to the $JLS$ basis.

\subsection{Partial Wave Amplitudes}
\label{subsec:PWA}

The Bethe-Salpeter type integral equation given in Eq.~(\ref{eq:Bethe-Salpeter}) is part of the production amplitude given in Eq.~(\ref{eq:decayrate}). Analytic solutions of such dynamical equations are only known for driving terms consisting of contact interactions, see e.g., Refs.~\cite{Mai:2013cka,Bruns:2010sv}. Therefore, the equation will be solved numerically here, by discretizing momenta, and thus, transforming the integral equation into algebraic ones. 

The technical challenge in doing so is that the $B$-term, and with it $T$, depends on the in- and outgoing spectator three-momenta, ${\bm p}$ and ${\bm q}_1$, making numerical inversions computationally demanding. However, only one of the terms of the partial wave decomposition in Eq.~\eqref{eq:pwa-of-T} enters the production amplitude Eq.~\eqref{eq:decayrate} for the quantum numbers of the $a_1(1260)$-resonance, i.e., the one for $J=1$.

\begin{figure}[t]
\centering
	\[
	\scalebox{0.99}
	{
		\begin{tikzpicture}[baseline=-0.13cm]
		\begin{feynman}[baseline=(a0)]
		\begin{feynhand}
		\vertex [] 	                      (a0){};
		\vertex [below=1cm of a0] 	      (ad){};
		\vertex [above=1cm of a0] 	      (au){};
		
		\vertex [right=2cm of a0]         (b0){};
		\vertex [below=1cm of b0] 	      (bd){};
		\vertex [above=1cm of b0] 	      (bu){};
		\propag[double, very thick] (au) to (bd);
		\propag[very thick]         (ad) to (bu);
		
		\node[draw,scale=0.9,diamond,fill=orange!50] at (1,0)  {\bf$C$};
		\end{feynhand}
		\end{feynman}
		\end{tikzpicture}
	}
	\scalebox{1.2}{$\boldsymbol{\Longrightarrow}$}
		\scalebox{0.99}
	{
		\begin{tikzpicture}[baseline=-0.13cm]
	\begin{feynman}[baseline=(a0)]
	\begin{feynhand}
	\vertex [] 	                      (a0){};
	\vertex [below=1cm of a0] 	      (ad){};
	\vertex [above=1cm of a0] 	      (au){};
	
		\vertex [right=3.5cm of a0]     (b0){};
	
	\vertex [right=0.925cm of a0]		(a1);
		\vertex [above=.07cm of a1]		(a2);
		\vertex [below=.07cm of a1]		(a3);
	\vertex [left=1cm of b0]		        (b1);
		\vertex [above=.07cm of b1]			(b2);
		\vertex [below=.07cm of b1]			(b3);

	\vertex [below=1cm of b0] 	      (bd){};
	\vertex [above=1cm of b0] 	      (bu){};

	\propag[double, very thick] (au) to (a1);
	\propag[very thick]         (ad) to (a1);
	
	\propag[very thick] (a1) to [edge label={$a_1(1260)$}] (b1);
	\propag[very thick] (a2) to (b2);
	\propag[very thick] (a3) to (b3);
	
	\propag[very thick] (bu) to (b1);
	\propag[double, very thick]         (bd) to (b1);
	
	\filldraw[fill=Black, line width=0.2mm] (0.8,-0.125) rectangle (1.05,0.125);
	\filldraw[fill=Black, line width=0.2mm] (2.4,-0.125) rectangle (2.65,0.125);
	\end{feynhand}
	\end{feynman}
	\end{tikzpicture} 
	}
	\]
\caption{
        Graphical representation of the isobar-spectator contact term $C$ via a propagation of a bare $a_1(1260)$. The coupling to $\rho\pi$ states is chosen consistently to that appearing in Fig.~\ref{fig:decayprocess}. See Eq.~\eqref{eq:contactterm} and text below it for more details.}
\label{fig:contactterm}
\end{figure}

The implementation of physical constraints on the three-body force induced term $C$ in Eq.~\eqref{eq:Bethe-Salpeter} can be made in the $JLS$ basis. The latter basis encodes the total-, relative (between isobar and spectator) and intrinsic (spin) angular momenta, respectively. In this basis we assume a general form of $C_{L'L}$ accounting for the partial wave dependence. Expressing the contact-term in terms of a propagating $a_1(1260)$ connecting initial and final isobar and spectator, as shown in Fig. \ref{fig:contactterm}, one may write
\begin{align}
\label{eq:contactterm}
C_{L'L}(p,q_1)=
\left(\frac{p}{m_\pi}\right)^{L'}
\left(\frac{q_1}{m_\pi}\right)^{L}
\frac{m^2_\pi g_{fL}g_{fL'}H(p)H(q_1) }{s-m^2_{\rm fit}} \,.
\end{align}
Here $g_{fL}$, $g_{fL'}$ are the bare couplings that characterize the strength of the decay vertex  and $m_{\rm fit}$ is the bare mass of the $a_1(1260)$. These parameters are fixed to reproduce physical data on the $a_1(1260)$ lineshape.  The form-factor $H$ is discussed in Appendix~\ref{app:formfac}. Note that factors of $m_\pi$ are included such that the above contact term is dimensionless. 

The term $T_{\lambda\lambda'}$ in Eq.~(\ref{eq:decayrate}) describes the isobar-spectator interaction symbolized in Fig.~\ref{fig:T-matrix}. For the purpose of the present paper only the part projected to total angular momentum $J=1$, the quantum number of $a_1(1260)$, is required. Taking into account the azimuthal symmetry the plane wave isobar-spectator amplitude is related to the partial wave amplitudes as 
\begin{align}
\label{eq:pwa-of-T}
A_{\lambda'\lambda}({\bm p}, {\bm q}_1)=\sum_J \frac{2J+1}{4\pi}\,d^J_{\lambda'\lambda}(z)A_{\lambda'\lambda}^J (p,q_1) \,,
\end{align}
where $A\in\{T,B\}$  and $d^J_{\lambda\lambda'}(\cos{\theta})$ denotes the small Wigner-d function.

In the $JLS$ basis and with in- and outgoing orbital angular momenta, $L'$ and $L$, Eq.~(\ref{eq:Bethe-Salpeter}) becomes 
\begin{align}
\label{eq:Bethe-SalpeterPartial}
T^J_{L'L}(p,q_1)&=\left(B^J_{L'L}(p,q_1)+C_{L'L}(p,q_1)\right)+
\\
&\sum_{L''}\int\limits_0^\infty \frac{dl\, l^2}{(2\pi)^3 2E_l} \left(B^J_{L'L''}(p,l)+C_{L'L''}(p,l)\right)\nonumber\\
&\hphantom{\sum_{L''}\int\limits_0^\infty \frac{dl\, l^2}{(2\pi)^3 2E_l} }\quad\qquad\times \tau(\sigma(l)) T^J_{L''L}(l,q_1)\nonumber
\,.
\end{align}
The $a_1(1260)$ is constrained by parity conservation and  conservation of angular momentum to decay into $\rho\pi$ with angular momentum $L=0$ or $L=2$. Thus, we work in a basis in which the $T$-matrix can have two $JLS$ states, namely $121$ and $101$ for both in- and outgoing states.

To obtain the $B$-term in the partial wave ($JLS$) basis we employ a two-step procedure. First, from the $B$-term \eqref{eq:BmatrixPlane} in helicity basis the relevant partial wave is extracted, exploiting orthonormality of Wigner-d functions,
\begin{align}
\label{eq:BmatrixPartial}
B^J_{\lambda'\lambda}(p,q_1)=
2\pi\int\limits_{-1}^{+1} dz \, d^J_{\lambda'\lambda}(z) B_{\lambda'\lambda}(&{\bm p},{\bm q}_1) \,.
\end{align}
Then, the expression in partial wave basis is obtained from a linear transformation
\begin{align}
\label{eq:transformation}
B^J_{L'L}(p,q_1)=U_{L'\lambda'}B^J_{\lambda'\lambda}(p,q_1) U^T_{L\lambda}\,,
\end{align}
where the superscript $(..)^{T}$ denotes the transposition operation. The transformation matrix $U$ is given by
\begin{align}
U_{L\lambda }=\sqrt{\frac{2L+1}{2J+1}}(L01\lambda|J\lambda)(1\lambda00|1\lambda) \,,
\end{align}
expressed in usual Clebsch-Gordan coefficients~\cite{Chung:1971ri}, while the summation over identical indices in Eq.~(\ref{eq:transformation}) is understood.

The above equations~\eqref{eq:Bethe-SalpeterPartial},
 \eqref{eq:BmatrixPartial} and \eqref{eq:contactterm} complete the main part of the final-state interaction to be solved numerically below. Ultimately, the solution of Eq.~\eqref{eq:Bethe-SalpeterPartial} in the $JLS$ basis needs to enter the decay amplitude~\eqref{eq:decayrate} as follows. First, the $a_1(1260)\rightarrow \rho \pi$ vertex, $D$ is modeled to include the correct spectator momentum dependence for each partial wave, i.e., $D_L(p)\sim p^L$. Furthermore, 
the production process itself may contain first-order singularities. The necessity of this contribution is explained at the end of Sec.~\ref{subsec:Inversion}. Overall, the vertex is parametrized as
\begin{align}
\label{eq:firstdecayvertexprime}
D_{L'}(p)=D_{fL'}
H(p)&
\left(\frac{p}{m_\pi}\right)^{L'}+ \\
\nonumber
&\qquad\quad
\frac{m^2_\pi g_{fL'} D_{\tilde{f}} H(p)}{s-m^2_{\rm fit}}\left(\frac{p}{m_\pi}\right)^{L'} \,,
\end{align}
where $D_{fL'}$ for $L'=0,2$ and $D_{\tilde{f}}$ are free parameters that are fit to the lineshape data as described below. The parametrization of Eq.~(\ref{eq:firstdecayvertexprime})
is similar to the one used in dynamical coupled-channel approaches for the photo-excitation of resonances, see, e.g., Ref.~\cite{Ronchen:2014cna}, where it allows to excite resonances and background independently without spoiling Watson's theorem.

Including the contact interactions $C$ and $D$ as in Eq.\eqref{eq:contactterm} and \eqref{eq:firstdecayvertexprime}, respectively, we construct the decay amplitude in the $JLS$ basis. We use the breve symbol on $\breve{\Gamma}$ to denote the inclusion of all terms represented diagrammatically in Fig.~\ref{fig:decayprocess} except  the final vertex ($v$), i.e., the diagram to the right of the parentheses. It can be separated into the contribution from the connected and the disconnected part with $\breve{\Gamma}=\breve{\Gamma}^C+\breve{\Gamma}^D$, where

\begin{align}
\label{eq:decayamplitudebreve-JLS}
\breve{\Gamma}_{L} (q_1)
&=D_{L}(q_1)\tau(\sigma(q_1)) \\
&+
\int \frac{dp \, p^2}{(2\pi)^3 2 E_p}
D_{L'}(p)
\tau(\sigma(p))
T^{J}_{L'L}(p,q_1)
\tau(\sigma(q_1))\, \nonumber.
\end{align}
with sum over $L'$.

Finally, this term is related to the amplitude in Eq.~(\ref{eq:decayrate}) by transforming into the helicity basis and multiplying the final $\rho\pi\pi$ vertex with
\begin{align}
\Gamma_{\Lambda\lambda} ({\bm q}_1,{\bm q}_2,{\bm q}_3)=
\sqrt{\frac{3}{4\pi}}&{\mathfrak D}^{1*}_{\Lambda\lambda}(\phi_1,\theta_1,-\phi_1)\\
&~~~\times \breve\Gamma_{L}(q_1)U_{L\lambda}v_{\lambda}({\bm q}_2,{\bm q}_3)\nonumber
\end{align}
where it is again summed over angular momentum $L$ and $\mathfrak{D}^{J}_{\Lambda\lambda}(\phi_1,\theta_1,-\phi_1)$ denotes   the capital Wigner-D function  with angles $\theta_1$ and $\phi_1$ giving the polar and azimuthal angles of $\bm{q}_1$. We use the convention of Jacob and Wick~\cite{Jacob:1959at} for the arguments of the function, $\mathfrak{D}^{J}_{\Lambda\lambda}(\phi, \theta, -\phi)$, rather than the alternative convention  $\mathfrak{D}^{J}_{\Lambda\lambda}(\phi, \theta, 0)$. The latter convention implies an additional phase-factor.

Note also that, while the small Wigner-d function may be used for the evaluation of the angular integration appearing in the partial-wave decomposition of Eq.~\eqref{eq:pwa-of-T}, one must use the capital Wigner-D functions and their $\phi$ dependence for the back-transformation to plain waves because in the symmetrized decay $a_1\to \pi^+\pi^-\pi^-$ of  Eq.~\eqref{eq:decayrate} the final $\rho^0$ isobars can be produced in different directions. 
In the next section we describe how the observables can be obtained from the production amplitude given above.

\subsection{Relation to Observables}
\label{subsec:Observables}

After having specified the analytic expressions leading to the production amplitude $\Gamma$ we will demonstrate in the following how it can be related to  three-body observables. In particular, this will allow us later to fix free parameters of the framework, i.e., $g_{fL},D_{fL'},m_{\rm fit}$, and $D_{\tilde{f}}$. Recall that parameters $g_1,\Lambda,m_\rho$ are already fixed to the two-body data -- the $\pi\pi$ phase-shifts in the isovector channel.
The three-body observables considered in this paper are  Dalitz plots, projected Dalitz plots and the lineshape.

The lineshape is a one dimensional scalar function of the total three-body energy $\sqrt{s}$. It is given by integrating over all three pion four-momenta in the final state
\begin{align}
\label{eq:lineshape}
\mathcal{L}(\sqrt{s})&=\frac{1}{\sqrt{s}}
\int \frac{d^3{\bm q}_1}{(2\pi)^3}
\frac{d^3{\bm q}_2}{(2\pi)^3}
\frac{d^3{\bm q}_3}{(2\pi)^3}
\frac{1}{2E_{q_1}2E_{q_2}2E_{q_3}}\\
&\quad
\times(2\pi)^4\delta^4(P_3-q_1-q_2-q_3)
\overline{|\Gamma(\boldsymbol{q}_1,\boldsymbol{q}_2,\boldsymbol{q}_3)|}^2\,.
\nonumber
\end{align}
Here the bar over the production amplitude $\Gamma$ denotes the usual summation over helicity indices, i.e., ${\overline{|\Gamma|}^2\equiv\nicefrac{1}{3}\sum_\Lambda\left|\sum_\lambda\Gamma_{\Lambda\lambda}\right|^2}$.

The Dalitz plots are calculated in a similar fashion, taking, however, the phase-space integral for fixed invariant masses $\sigma_{23}$ and $\sigma_{13}$ made up of the two outgoing pions given in the subscript,
with $3$ labeling the $\pi^+$ and $1,2$ the two $\pi^-$,
\begin{align}
\label{eq:DalitzPlot}
\mathcal{D}(\sqrt{s},\sigma_{23},\sigma_{13})=&
\frac{1}{(2\pi)^5}\frac{1}{32\sqrt{s}^3 }\int d \Omega_{q_1}
d\phi_{12}\\
\nonumber
&\qquad\quad\qquad\times\overline{|\Gamma (\sigma_{23},\sigma_{13},\Omega_{q_1},\phi_{12})|}^2\,.
\end{align}
Note that the delta function, ${\delta^4(P-q_1-q_2-q_3)}$ for energy-momentum conservation has been evaluated, which accounts for the elimination of the differential $d^3\bm{q}_3$ and one of the angles. The latter angle is chosen to be the angle between $\bm{q_1}$ and $\bm{q_2}$, denoted by $\theta_{12}$. The azimuthal angle between the same momenta, denoted by $\phi_{12}$, remains in the integration.

Finally, taking the Dalitz distribution of Eq.~\eqref{eq:DalitzPlot} and integrating over one of the invariant masses, the projected Dalitz plot is obtained with
\begin{align}
\label{eq:ProjectedDalitzPlot}
\mathcal{D}_p(\sqrt{s},\sigma_{23})= & 
\frac{1}{(2\pi)^5}\frac{1}{32\sqrt{s}^3 }
\\ & \nonumber
 \int d \Omega_{q_1} 
d\phi_{12} d \sigma_{13} \overline{|\Gamma (\sigma_{23},\sigma_{13},\Omega_{q_1},\phi_{12})|}^2\,.
\end{align}

The numerical treatment for these integrals  is discussed in Sec.~\ref{subsec:MonteCarlo} and results are shown in Sec.~\ref{subsec:numresults}.

\section{Numerical implementation}
\label{subsec:implementation}

\begin{figure}[t]
\centering
\includegraphics[width=1\linewidth]{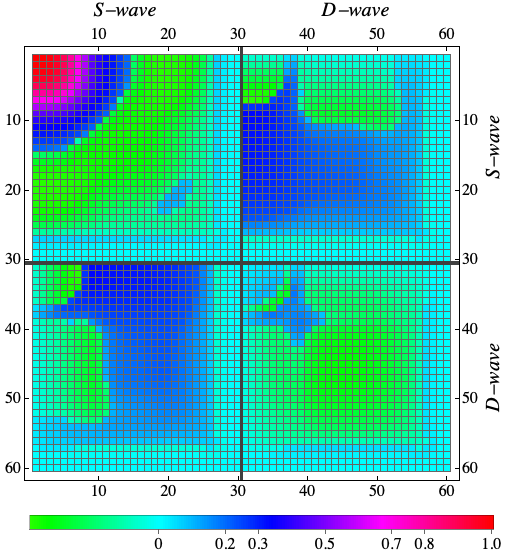}
\caption{Discretized version of $\Re B^J_{p,q_1}$ from Eq.~\eqref{eq:B-matrix}. X- and Y-axes correspond to the in- and outgoing combined momentum and angular momentum index as explained in Sec.~\ref{subsec:Inversion}. Note that the momenta are not uniformly distributed.}
\label{fig:MP-plot}
\end{figure}

\subsection{Inversion of the Bethe-Salpeter Equation}
\label{subsec:Inversion}

The Bethe-Salpeter type integral equation given in the plane wave basis in Eq.~\eqref{eq:Bethe-Salpeter} and the partial wave basis in Eq.~\eqref{eq:Bethe-SalpeterPartial} must be solved in order to calculate the observables discussed in Sec.~\ref{subsec:Observables}. This is done numerically by discretizing the analytic functions of in- and outgoing momenta in the $T$, $\tau$, $B$- and $C$-terms. 

We start the discretization by introducing indices $p$ and $q_1$ (not to be confused with fourmomenta). They both sample the region $(0,\infty)$ on a suitably mapped Gaussian quadrature, consecutively for $L=0$ and $L=2$, see Fig.~\ref{fig:MP-plot} for an example. 

This mapping allows one to write the terms $B_{LL'}^J(p,q_1)$ and $T_{LL'}^J(p,q_1)$ in a compact matrix form $B_{pq_1}^J$ and $T_{pq_1}^J$, respectively, preserving the dependence on $L$ and $L'$. In this notation Eq.~\eqref{eq:Bethe-SalpeterPartial} reads 
\begin{align}
\label{eq:B-matrix}
T^{J}_{pq_1}=\left( B^{J}_{pq_1}+C_{pq_1}\right)+\left(B^{J}_{pl}+C_{pl}\right)\tilde{\tau}_l T^{J}_{l q_1} \,,
\end{align}
where the summation over equal indices is understood. To simplify the notation, factors of $l^2/((2\pi)^3 2E_l)$ and Gaussian integration weights are absorbed into the isobar propagator, indicated by the diagonal expression $\tilde{\tau}$. The solution of this equation  reads
\begin{align}
T^{J}_{pq_1}=\left(\left[\mathds{1}-
\left( B^{J}+C\right)\tilde\tau\right]^{-1}\right)_{pl}\left( B^{J}_{lq_1}+C_{lq_1}\right) \,,
\label{eq:Tmatrixnumerical}
\end{align}
with $\mathds{1}$ denoting the unit matrix in the above defined space, i.e., the direct product of momenta and angular momenta. We introduce a discretized version of $\breve{\Gamma}$ following Eq.~\eqref{eq:decayamplitudebreve-JLS}
\begin{align}
\label{eq:decayratenumerical}
\Breve{\Gamma}_{q_1}
= \big(
D_{q_1}\tau_{q_1}+ 
D_{p}
\tilde\tau_p
T^{J}_{pq_1}
\tau_{q_1}
\big)\,.
\end{align}
The $L$ dependence is implicit in the $q_1$ index.

Before proceeding to further details of the numerical implementation we return to the discussion of  the $D$-term in Eq.~\eqref{eq:firstdecayvertexprime} and why it is important to include the same singularity in the $D$-term as in the $C$-term. Consider first the three-body contact term $C_{pq_1}$~that is the discretized version of Eq.~\eqref{eq:contactterm}, which diverges for $\sqrt{s}\to m_{\rm fit}$. Following Eq.~\eqref{eq:Tmatrixnumerical}, in this limit $T^{J}_{pq_1}\to-1/\tilde\tau_{q_1}\delta_{pq_1}$. Subsequently, this yields for the production amplitude Eq.~\eqref{eq:decayratenumerical}, $\Breve{\Gamma}_{q_1}\to (D_{q_1}\tau_{q_1}-D_{p}\delta_{pq_1}\tau_{q_1}) =0$. 
There is, however, no physical reason for the amplitude to vanish at this specific point. Adding a contribution, singular at $s=m_{\rm fit}^2$, such as the second term in Eq.~\eqref{eq:firstdecayvertexprime}, solves this issue.

\begin{figure}[t]
\centering
\includegraphics[width=1\linewidth]{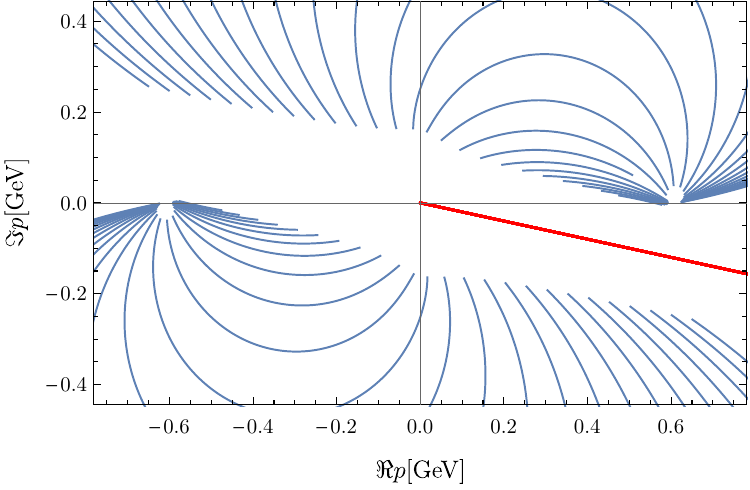}
\caption{Illustration of the singularity structure of the $B$-term. The red and blue lines mark the integration contour for the momentum $p$ and the moving singularities in the $B$-term, respectively. Each of the blue lines is fixed by one complex $q_1$ for various values $\cos\theta\in(-1,1)$. The former values are representative for the problem at hand, while $\sqrt{s}=1.260$~GeV. Note that changing the integration contour affects both, the red and blue lines.}
\label{fig:IntegrationContour}
\end{figure}

\subsection{Integration Contour}
\label{sec:integration_contour}

While the quantities of interest  $\breve{\Gamma}_{L}$  and $T^{J}_{pq_1}$ read quite straightforwardly in Eqs.~\eqref{eq:Tmatrixnumerical} and \eqref{eq:decayratenumerical}, there is a complication hidden in the analytic structure of its building blocks. The main issue in this respect deals with the cuts of the $B$-matrix and self-energy given by Eq.~\eqref{eq:BmatrixPartial} and Eq.~\eqref{eq:selfenergy}. Utilizing analyticity, we avoid these cuts by deforming the integration contour, i.e., sampling all momentum magnitudes on a line in the complex plane as, e.g., $\Im k/\Re k=-\alpha$~\cite{Hetherington:1965zza}. In order to solve Eq. (\ref{eq:B-matrix}) with respect to $T$, one has to choose both momenta $p$ and $q_1$ along the same deformed integration contour. This contour, used for the momentum integration in Eq. \eqref{eq:decayamplitudebreve-JLS}, is depicted by the red line in Fig. \ref{fig:IntegrationContour}. The figure also illustrates the singularities of the $B$-matrix (blue lines) which are circumvented by the deformed integration contour. 
The form-factors regularizing the $T$ matrix, see Eqs.~\eqref{eq:finalvertex} and \eqref{eq:contactterm}, also develops similar poles. We have made sure for all
moving singularities appearing in the problem that they do not cross the integration contour, and that these singularities respect the $+i\epsilon$ prescription of Eq.~\eqref{eq:BmatrixPlane} for energies $\sqrt{s}$ above threshold.

The fact that the integration contour avoids poles guaranties analyticity of the three-body scattering amplitude $T$. In principle, the integration contour is not unique and other choices are possible, but there are certain constraints on it. On the one hand, $\alpha$ needs to be small enough, such that the analytic extrapolation (see below) is a good approximation of $\breve{\Gamma}_{q_1}$ for real $q_1$. On the other hand, it needs to be large enough to maintain numerical stability. After extensive exploration, we choose $\alpha=0.2$ fulfilling the above constraints. Recall that the pole-structure of the self-energy in Eq.~\eqref{eq:selfenergy} is not as intricate. For simplicity, we choose the same integration contour for it as for the $B$-term. This also ensures that the correct Riemann sheet for the self-energy is picked.

The integration for the self-energy and the integration over the internal spectator momentum, $p$ in Eq.~\eqref{eq:decayrate}, are taken over all momentum. Thus, integration on our chosen contour is identical to integration along the real axis. This is ensured by the form-factors which cause the integrand to vanish at large momenta. However, the outgoing spectator momentum, $q_1$ is real. Because $q_1$ can be observed, $\breve{\Gamma}$ is integrated to the physical limit determined by $\sqrt{s}$ in the calculation of observables. In order to relate the results for momenta on the chosen contour to those on the real axis we fit a Padé-approximant to the numerically obtained values of $\breve{\Gamma}$ at complex $q_1$. We then extrapolate this function to the real axis.
Note that this is possible because there are no further non-analyticities between the complex contour and the real $q_1$ axis.
Furthermore, the incoming momentum $p$ of the $T$-matrix is always complex and we never encounter the situation with both $p$ and $q_1$ real, which could induce further singularities into the $B$-matrix~\cite{Hetherington:1965zza}.

To parametrize $\breve{\Gamma}$, its known asymptotic ($q_1\rightarrow 0$, $q_1\rightarrow\infty$) behaviour can be explicitly incorporated into the Pade-approximant as
\begin{align}
\label{eq:PadeApproximant}
\breve{\Gamma}_{L}(q_1)=\left(\frac{q_1}{m_\pi}\right)^L  H(q_1)\frac{\sum_{j=0}^m a^{L}_{j}(q_1/m_\pi)^j}{\sum_{k=0}^nb^{L}_{k}(q_1/m_\pi)^k} \ ,
\end{align}
which approximates the right-hand side of Eq.~\eqref{eq:decayamplitudebreve-JLS}.
The complex coefficients, $a_j^{L}$ and $b_k^{L}$ are fit to the values of $\breve{\Gamma}_{q_1}$ of Eq.~\eqref{eq:decayratenumerical}, that is calculated only for discrete, complex $q_1$. The Padé-approximant is then extrapolated to real values of $q_1$.
The inclusion of the asymptotic behavior ensures that one only needs a small number of free parameters to accurately fit $\breve{\Gamma}_{q_1}$.

We have extensively checked the validity of this procedure, e.g., by ensuring that the extrapolated function is independent of the choice for $\alpha$ with sufficient precision.
The numerical precision also depends on the the number of Gauss points used for the discretization of Eq.~\eqref{eq:decayratenumerical}. Tests have shown that with 30 Gauss points for each partial wave one achieves a discretization error of similar size as the extrapolation error, which is the number employed in this study.

\subsection{Monte Carlo Integration}
\label{subsec:MonteCarlo}

After having calculated the production amplitude $\Gamma$ we need to perform the phase-space integrations of Eqs.~\eqref{eq:lineshape}, \eqref{eq:DalitzPlot} and~\eqref{eq:ProjectedDalitzPlot}, leading to the lineshape and Dalitz plots. We do this numerically using Monte Carlo (MC) integration which involves summation over the function for randomly generated values of relevant kinematic  variables.

The number of integration variables in Eq.~\eqref{eq:lineshape} is reduced from nine to five, accounting for conservation of 4-momentum encoded in the delta function. In doing so, we replace the differential from $d^3\bm{q}_1 d^3\bm{q}_2 d^3\bm{q}_3$ to $dq_1 dq_2 d\phi_1 d \cos{\theta_1} d\phi_{12}$. The angles $\phi_1$ and $\theta_1$ are the angles, which $q_1$ forms with the $z$-direction. The angle $\phi_{12}$ is the azimuthal angle between $q_1$ and $q_2$. 

The integral is then calculated using the Monte Carlo method, implemented similarly to Ref.~\cite{Doring:2010fw}. This involves $N$ randomly generated sets each consisting in the corresponding integration variables. Each value is denoted with a subscript, for example, the $i$-th value of $\phi_{12}$ is denoted $\phi_{12i}$. Furthermore, the allowed values for $q_1$ and $q_2$ must satisfy the condition $|\cos{\theta_{12}}|<1$, where $\cos{\theta_{12}}$ is fixed by the delta function to be
\begin{align}
\label{eq:costheta}
\cos\theta_{12}(q_1,q_2)=\frac{(\sqrt{s}-E_{q_1}-E_{q_2})^2-m_\pi^2-\bm{q}_1^2-\bm{q}_2^2}{2q_1q_2}\,.
\end{align}
The explicit formula for the MC integration of the lineshape reads
\begin{align}
\label{eq:LineShapeNumerical}
\mathcal{L}(\sqrt{s})=&\frac{1}{(2\pi)^4}\frac{P_L}{N} \frac{1}{8\sqrt{s} } \\
\nonumber
&\times\sum_{i=1}^N \frac{q_{1i} q_{2i}}{E_{q_{1i}}E_{q_{2i}}}\overline{|\Gamma(q_{1i},q_{2i},\cos{\theta_{1i}},\phi_{1i},\phi_{12i})|}^2 \,,
\end{align}
where $N$ is the number of MC points and $P_s$ is the phase space factor, given by integration over the volume. For the lineshape it is
\begin{align}
P_L=\int dq_1 dq_2 d \cos\theta_{1}\,d\phi_1 d\phi_{12} \,\Theta(1-\cos^2{\theta_{12}(q_1,q_2)}) \,,
\end{align}
where $\Theta$ is the Heaviside function.
In a similar way the Dalitz plot 
\begin{align}
\label{eq:DalitzNumerical}
\mathcal{D}(\sqrt{s},\sigma_{23},\sigma_{13}) = &\frac{1}{(2\pi)^4} \frac{P_D}{N}\frac{1}{32\sqrt{s}^3 }
\\ & \nonumber
\times
 \sum_{i}  \overline{|\Gamma(\sigma_{23},\sigma_{13},\cos{\theta_{1i}},\phi_{1i},\phi_{12i})|}^2 
\end{align}
and the projected Dalitz plot
\begin{align}
\label{eq:ProjectedDalitzNumerical}
\mathcal{D}_P(\sqrt{s},\sigma_{23}) =& \frac{1}{(2\pi)^4} \frac{P_P}{N} \frac{1}{32\sqrt{s}^3 }
\\ & \nonumber
\times
 \sum_{i}  \overline{|\Gamma(\sigma_{23},\sigma_{13i},\cos{\theta_{1i}},\phi_{1i},\phi_{12i})|}^2 
\end{align}
are calculated. The phase space factor for the Dalitz plot and projected Dalitz plots are $
P_D=\int  d \cos\theta_{1}d\phi_1 d\phi_{12}  $ and $
P_P=\int d\sigma_{13} d\cos\theta_{1} \,d\phi_1 d\phi_{12} \,\Theta(1-\cos^2{\theta_{12}(\sigma_{13},\sigma_{23})}) $ respectively.

\begin{figure}[t]
\includegraphics[width=1.\linewidth]{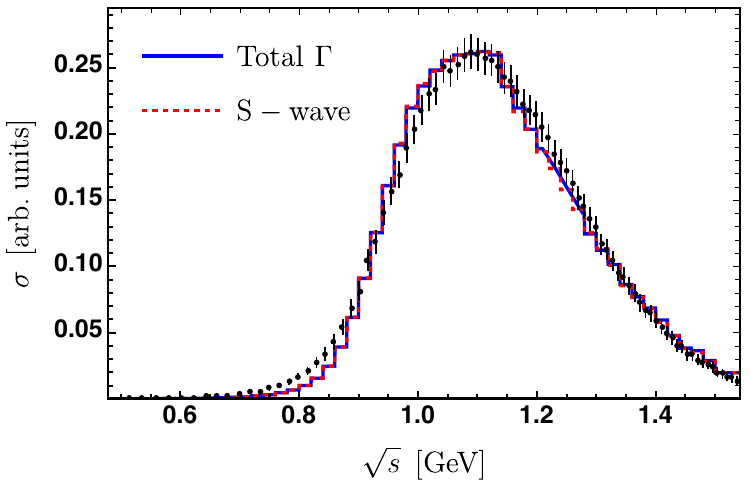}
\caption{Three-pion lineshape from the decay $\tau^-\rightarrow\pi^-\pi^-\pi^+\nu_\tau$ (data from Ref.~\cite{Schael:2005am}). The solid blue line shows the fit result. We also include the component of the fit that comes only from the $S$-wave (red line). As expected, the $S$-wave channel dominates. We also see that the $D$-wave provides a larger contribution at higher values of $\sqrt{s}$.
}
\label{fig:LineShape}
\end{figure}

\begin{figure*}[t]
\centering
\includegraphics[width=1.05\linewidth, 
trim=0cm 0cm 0cm 0cm, 
clip ]{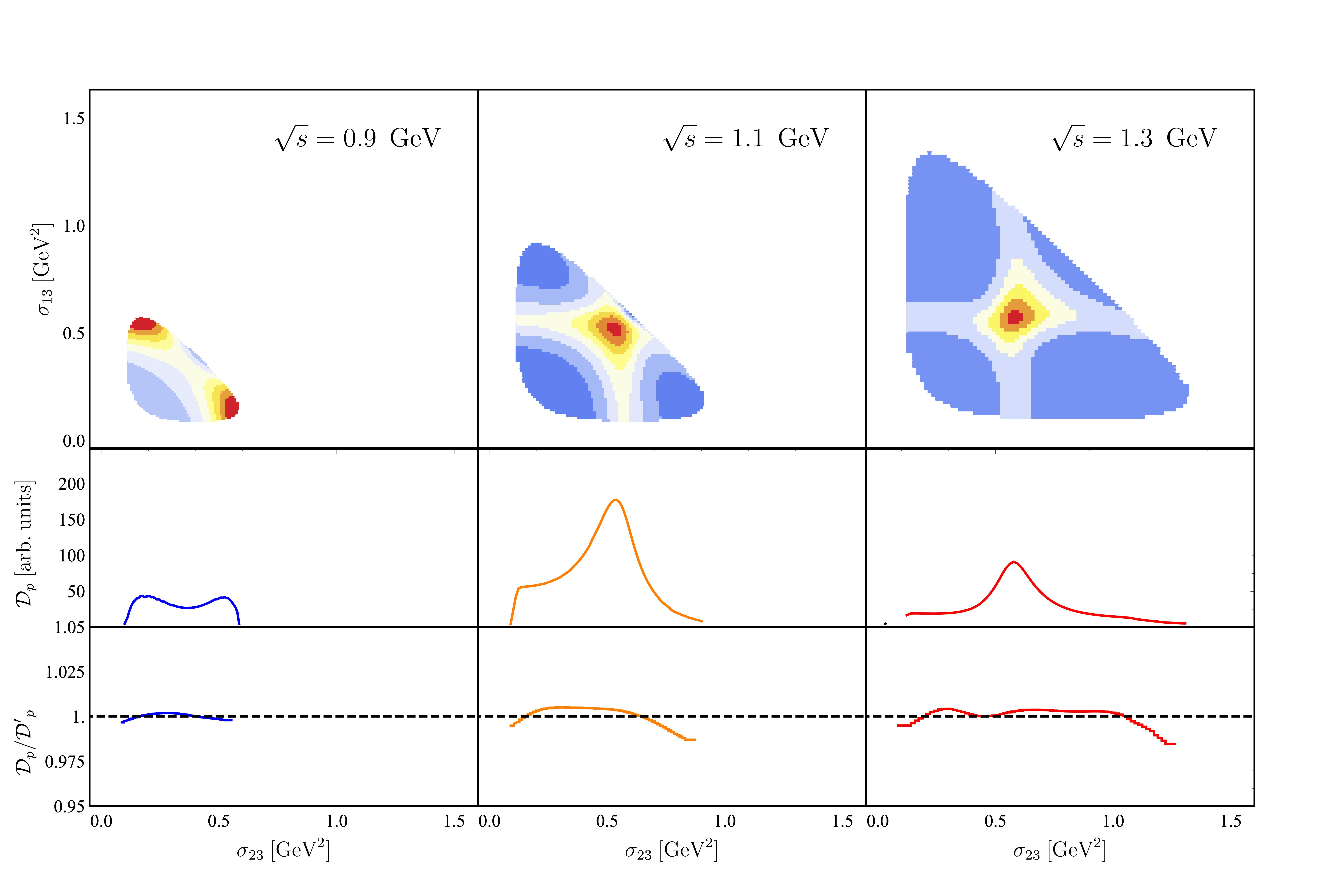}
\caption{Upper Row: Dalitz plots for various three-body invariant masses. Each Dalitz plot has been re-scaled separately to emphasize qualitative features. The blue/red colors represent higher/lower values, respectively. The dark bands at $\sigma_{..}\sim0.59$~GeV${}^2$ are the outline of the $\rho$-meson.
Middle Row: Projected Dalitz plots. An effect of this integration is the kinematic reflection of the peak from the $\rho$ meson at a value of $\sigma_{13}$ lower than the mass of the $\rho$. 
Lower Row: The ratio between $\mathcal{D}_p$ and $\mathcal{D'}_p$, the equivalent projected Dalitz plots calculated without the rescattering terms normalized to the integral over $\mathcal{D}_p$.
\label{fig:DalitzPlots}
}
\end{figure*}

\section{Results}\label{subsec:numresults}

In the following we present the results of the approach in terms of Dalitz plot, projections thereof, and lineshape, i.e., $\mathcal{D}$, $\mathcal{D}_p$, and $\mathcal{L}$ calculated from Eqs.~\eqref{eq:DalitzNumerical}, \eqref{eq:ProjectedDalitzNumerical}, and~\eqref{eq:LineShapeNumerical}, respectively. 

The free parameters of the framework are $m_\rho$, $g_1$, $\Lambda$, $g_{fL}$, $D_{fL}$, $D_{\tilde f}$, $m_{\rm fit}$, where three former are fixed by a fit to the two-body data, i.e., isovector $\pi\pi$ phase-shifts. They read: $m_\rho=1.82$~GeV, $g_1=7.23$, and $\Lambda=1.58$. As discussed below, there is not much sense in determining statistical uncertainties for these quantities for the present purpose. Note that $\Lambda$, appearing in form-factor $F$ in the $\rho\pi\pi$ vertex of Eq.~\eqref{eq:finalvertex}, regularizes not only the isovector two-body amplitude but also the three-body amplitude through the appearance in the $B$-term of Eq.~\eqref{eq:BmatrixPlane}. This induces the usual and unavoidable regulator dependence in the three-body amplitude which is absorbed in the contact term \eqref{eq:contactterm}, that by itself contains a form-factor to render the three-body equation well-defined.

The remaining six parameters (note the index $L=0,2$ above) are determined from a fit to the lines-shape from the ALEPH~\cite{Schael:2005am} experiment. The best fit parameters read:  $D_{f0}=1.4\times 10^{-4}~\rm{GeV}^{-1/2}$, $D_{f2}=2.7\times10^{-5}~\rm{GeV}^{-1/2}$ for the non-resonant $\rho\pi$ production vertices of Eq.~\eqref{eq:firstdecayvertexprime}, $D_{\tilde{f}}=2.7\times10^{-7}~\rm{GeV}^{-1/2}$ for the strength of the bare $a_1$ component in the production process in Eq.~\eqref{eq:firstdecayvertexprime}, $g_{f0}=2.8\times10^{3}$, $g_{f2}=1.09$,
and $m_{\rm fit}=1.232$~GeV
for the bare $a_1$ decays into $S$- and $D$-wave $\rho\pi$ and the $a_1$ bare mass of Eqs.~\eqref{eq:contactterm} and \eqref{eq:firstdecayvertexprime}, respectively.  The result of the fit is presented in Fig.~\ref{fig:LineShape} together with the experimental data.

According to the PDG~\cite{Tanabashi:2018oca}, the $a_1(1260)$ mostly decays into the $\rho\pi$ and  $\sigma\pi$ channels. The $\sigma\pi$ branching ratio is determined in Ref.~\cite{Asner:1999kj} to be $\Gamma(\sigma\pi)/\Gamma_{total}=0.1876$, whereas  relative to the $\rho\pi$ channel it is quoted in Ref.~\cite{Salvini:2004gz} as $\Gamma(\sigma\pi)/\Gamma(\rho\pi)=0.06$. Both, these results show that the $\rho\pi$ channel is largely dominant, but the $\sigma\pi$ channel is non-negligible. Since the present analysis does not include the $\sigma\pi$ channel, a $\chi^2$ close to 1  would only indicate that the free parameters are capable of accounting for effects that are probably not due to the physical processes they describe. Therefore, we do not preform a statistical analysis on our fit of the lineshape data and accept that the fit starts diverging from the lineshape data at at high energies as shown in Fig.~\ref{fig:LineShape}.
Furthermore, at higher energies the $\rho$ channel itself exhibits structures that are not included in our parametrization of the $\pi\pi$ amplitude, in terms of increasing inelasticty and resonances such as the $\rho(1450)$.

We can make a very rough estimate of the {$D$-wave} branching ratio (BR) from the lineshape at the resonance position $\sqrt{s}\approx 1.26$~GeV. We obtain ${\Gamma(\rho\pi)_D/\Gamma_{\rm{tot}}\approx 0.25}$~\%. This is comparable to the PDG value~\cite{Tanabashi:2018oca}, i.e.,  ${\Gamma(\rho\pi)_D/\Gamma_{\rm{tot}}= \left(1.30\pm 0.60\pm 0.22\right)\%}$.

Also, comparing to other determinations of BRs is difficult because in BRs for sequential decays, integrals are often performed over the mother and/or daughter resonances. This contribution depends on integration limits, the non-resonant background and the parametrization of the lineshapes of mother and daughter resonances. A unique and background-independent definition is provided by the residue at the pole~\cite{Doring:2009bi, Doring:2009yv}, which for unstable daughter resonances will be a function of the spectator momentum; the analytic continuation of the amplitude to the $a_1$ pole, as performed in Ref.~\cite{Mikhasenko:2018bzm}, is beyond the scope of this work because the moving three-body singularities appearing in our unitary formalism require special attention for complex energies~\cite{Doring:2009yv}.

In the upper row of Fig.~\ref{fig:DalitzPlots} we show the result of our calculation of the Dalitz plots using the parameters  from the fit to the lineshape. In those, we note the clear outline of the $\rho$-meson at $\sigma\sim0.59$~GeV${}^2$ for the plots with large enough available phase-space. The maxima of the plots lie at the intersection of the bands, meaning that the amplitudes exhibit a constructive interference. As such, this is consistent with the symmetry of the amplitude $\hat\Gamma_{\Lambda\lambda} (\boldsymbol{q}_1,\boldsymbol{q}_2,\boldsymbol{q}_3)$ under the exchange of the two $\pi^-$ due to Bose symmetry, see  Eq.~\eqref{eq:decayrate} and discussion below it.

The central row of Fig.~\ref{fig:DalitzPlots} shows the projected Dalitz plots. We once again see peaks in each plot at the $\rho$-mass as well as its kinematic reflections. To understand the importance of the connected contributions, we show in the bottom row of the figure the quotient of the full and disconnected contributions, cf., Eq.~\eqref{eq:decayrate}. We observe a contribution from  rescattering at the order of $\sim5\%$. We note that this contribution is small but of similar order as the branching ratios to some inelastic channels (not included in this analysis), see Ref.~\cite{Tanabashi:2018oca}. Thus, the incomplete inclusion of rescattering might lead to incorrect extraction of the resonance parameters in these channels.

\section{Summary}\label{sec:summary}

In this paper, we have adopted a relativistic unitary formalism for three-to-three scattering to address the decay of the $a_1(1260)$-resonance to three pions. The free parameters of the approach, related to the production mechanism, two-body sub-channel interaction and the three-body contact term were fixed to the two- and three-body data.

The key technical part of the approach is the relativistic Bethe-Salpeter type equation, which depends on the momenta of the in- and outgoing pions. First, this integral equation was reduced by a partial-wave decomposition. This coupled-channel equation has then been discretized for momenta on a complex contour, leading to an algebraic equation. Solutions of the latter have been extrapolated to real momenta, for which the amplitudes were then related to observables. These include lineshape, Dalitz plots, and projected Dalitz plots. 

Fitting the free parameters of the approach to the lineshape from the ALEPH experiment, we observe a dominant $S$-wave contribution. The Dalitz plots and projected Dalitz plots show the outline of the $\rho$-meson and its kinematic reflections. The rescattering effects are small but non-negligible for this particular system, and could affect the extraction of resonance parameters in a more complete description. The inclusion of $\sigma\pi$ and other channels, the direct analysis of Dalitz plots is work in progress. This would allow for an accurate determination of the pole position of the $a_1(1260)$ through analytic continuation to the complex energy plane.

\begin{acknowledgments}
The work is supported by the National Science Foundation CAREER grant PHY-1452055 and by the U.S. Department of Energy, Office of Science, Office of Nuclear Physics under contract no. DE-AC05-06OR23177 and grant no. DE-SC0016582. MM is grateful to the German Research Association (Project \# MA 7156/1), for the financial support during the development of theoretical foundations underlying this work. HA would like to thank the Avicenna-Studienwerk e.V. for their scholarship supporting the research semester at the GWU.
\end{acknowledgments}

\bibliography{ALL-REF.bib}

\appendix
\begin{onecolumngrid}

\subsection{Form-Factors}
\label{app:formfac}

The integral equations, appearing in the main text, are regularized using  form-factors. Apart from suppressing large spectator or loop momenta, the requirements on the form-factor in the $\rho\to\pi\pi$ decay $v$ are
\begin{enumerate} 
 \item
 \label{condition}
 As a consequence of three-body unitarity, the $\rho\pi\pi$ decay $v$ in the self-energy of Eq.~\eqref{eq:selfenergy}, in the $\pi$ exchange of Eq.~\eqref{eq:BmatrixPlane}, and in the final decay of Eq.~\eqref{eq:decayrate} must be consistent~\cite{Mai:2017vot}. In particular, the form-factor must be Lorentz invariant because it is evaluated in the three-body rest frame but also the isobar rest frames. 
 \item
 To preserve unitarity, the previous requirement must be fulfilled if the pions of the $\rho$ decay are on-shell, but may be dropped if they are off-shell.
 \item For the $B$-term, it cannot have poles on the chosen integration contour; this requirement needs to be fulfilled for all scattering angles, spectator momenta, and three-body energies $\sqrt{s}$. Also, it cannot contain poles in the region around the contour and the real-momentum axis, due to the extrapolation procedure to obtain the amplitude for real outgoing momenta from complex ones described in Sec.~\ref{sec:integration_contour}.
\end{enumerate}
Obviously, the above requirements do not fix the form-factor uniquely. Various analytic forms are studied and discussed in \cite{Sadasivan:2020}. For the $B$-term and the final $\rho\to\pi\pi$ decay, we choose a form-factor for Eq.~\eqref{eq:finalvertex} that reads 
\begin{align}
F(\sigma,Q^2)=\frac{\Lambda^4}{\Lambda^4+e^{1+(Q^2/4-(\sigma-4m_\pi^2))/(1 \rm{GeV}^2)}}\,
\label{eq:form1}
\end{align}
for $\sigma=(q_2+q_3)^2$, $Q^2=(q_2-q_3)^2$. As demanded by  condition \ref{condition} in the above list, a consistent choice needs to be made for the form-factor in the self-energy~\eqref{eq:selfenergy}. However, since the latter only includes vertices projected to $P$-wave, the form-factor will have a different momentum dependence. It reads
\begin{align}
\tilde{F}(\sigma,k)=\frac{\Lambda^4}{\Lambda^4+e^{1+((4(\sqrt{\sigma}-2E_k)^2-4k^2)/4-(\sigma-4m_\pi^2))/(1 \rm{GeV}^2)}}\,.
\label{eq:form2}
\end{align}
Note, that the term $(\sqrt{\sigma}-2E_k)^2$ vanishes for the on-shell momentum, $k=\sqrt{\sigma-m_\pi^2}/2$, which is the value leading to an imaginary part of the self-energy integral~\eqref{eq:selfenergy}. Therefore, we can multiply this term by an arbitrary factor and still satisfy condition \ref{condition} for the form-factors. We multiply it by a factor of four to allow the form-factor to suppress large momenta $k$ for the case when $\sigma(p)=0$. 

Another form-factor appears in the three-body term $C$ of Eq.~\eqref{eq:contactterm} and similarly in the $a_1(1260)\rightarrow \rho\pi$ vertex of Eq.~\eqref{eq:firstdecayvertexprime}. Both these contributions are real-valued and thus respect the unitarity requirement and furthermore enter only in a fixed (center-of-mass) reference frame. Hence, they do not have to follow the above constraints, and are chosen for simplicity as
\begin{align}
H(p)=\frac{\Lambda'^4}{\Lambda'^4+p^4}\,.
\label{eq:form3}
\end{align}
In principle this factor allows for an additional parameter $\Lambda'$. We varied it but found no significant improvement of the fits. Thus, we choose the cutoff $\Lambda'=1.0~\rm{GeV}$.

\subsection{  Polarization Vectors and Kinematics}
\label{app:Polarization}

The $\rho\to\pi\pi$ vertex in Eq.~\eqref{eq:finalvertex} includes the four-product $\epsilon_{\lambda}^\mu(\bm{q}_1)~(q_2-q_3)_\mu$ with $\lambda$ indicating the helicity of the isobar. The polarization vector or its complex conjugate is assigned to a given in- or outgoing isobar. Explicitly, the corresponding polarization vectors read,
\begin{align}
\label{eq:kinematics-INOUT}
   \epsilon_0&=\frac{1}{m}\begin{pmatrix} p \\  0 \\ 0 \\ E_p \end{pmatrix}\,,
   &\epsilon_{\pm 1}&=\frac{1}{\sqrt{2}}\begin{pmatrix} 0 \\  \mp 1 \\ -i \\ 0 \end{pmatrix}\,,
   &\epsilon^*_0&=\frac{1}{m}\begin{pmatrix} k \\  E_{q_1} \sin{\theta_1}\cos{\phi_1} \\ E_{q_1} \sin{\theta_1}\sin{\phi_1} \\ E_{q_1} \cos{\theta_1} \end{pmatrix}\,,
   &\epsilon^*_{\pm 1}&=\frac{1}{\sqrt{2}}\begin{pmatrix} 0 \\  \mp \cos{\theta_1}\cos{\phi_1}+i\sin{\phi_1} \\ -i\cos{\phi_1}\mp \cos{\theta_1}\sin{\phi_1} \\ \pm \sin{\theta_1} \end{pmatrix}
   \, ,
\end{align}
where the ingoing isobar with momentum $\bm p$  points in the $z$-direction, see, e.g., Eq.~\eqref{eq:Bethe-Salpeter}. In order to simplify the partial-wave projection of Eq.~\eqref{eq:BmatrixPartial}, azimuthal symmetry allows us to choose a reference frame, in which the outgoing isobar at momentum $\bm q_1$ lies in the $xz$-plane with scattering angle $\theta_1$ and $\phi_1=0$. 

The polarization vectors are also used for the calculation of the final vertex in Eq.~\eqref{eq:decayrate}. In this case, we work with a more general expression of the momenta in the phase space as defined in the integral of Eq.~\eqref{eq:DalitzPlot}.  We choose to define the angles of $\bm{q}_2$ relative to $\bm{q}_1$, i.e., the angle $\theta_{12}$ and the azimuthal angle $\phi_{12}$. Thus, one defines the explicit components of  $\bm{q}_2$ in terms of $\theta_{12}$, $\phi_{12}$, and the angles of the rotation of $\bm{q}_1$ from the $z$-axis in its actual direction. We introduce the rotation matrices $R_{\bm{q}_1}(\phi_{12})$ and $R_{\bm{q}_{1\bot}}(\theta_{12})$, where the subscript specifies the vector about which they are rotated and the argument gives the angle by which they are rotated. The vector $\bm{q}_{1\bot}$ is calculated by $R_{z}(\phi_{1})R_{y}(\theta_{1})
    R_{z}(-\phi_{1})\hat{y}$. Explicitly, one has
\begin{align}
\label{eq:kinematics-ROTATION}
   \bm{q}_1&=
  -q_1 \begin{pmatrix} 
   R_{z}(\phi_{1})R_{y}(\theta_{1})
    R_{z}(-\phi_{1})\hat{z}
   \end{pmatrix}\,,
   &\bm{q}_2&=
  q_2 \begin{pmatrix} 
   R_{\bm{q}_1}(\phi_{12})R_{\bm{q}_{1\bot}}(\theta_{12})
    R_{\bm{q}_1}(-\phi_{12})  \bm{\hat{q}}_1
   \end{pmatrix}\,,
\end{align}
whereas $\bm{q}_3$ is fixed to be $-\bm{q}_3=\bm{q}_1+\bm{q}_2$ in the overall center of mass frame. The labeling of momenta changes through the symmetrization of the two $\pi^-$ indicated in Eq.~\eqref{eq:decayrate}.
The components of the polarization vectors change depending on whether $q_1$ or $q_2$ is designated to be the spectator. If $q_1$ is the spectator, the polarization vectors will have the same components as the right two equations~\eqref{eq:kinematics-INOUT}, whereas if $q_2$ is the spectator, the equations read
\begin{align}
   \epsilon_{\cor{}{0}}=\frac{1}{m}\begin{pmatrix} k \\   R_{\bm{q}_1}(\phi_{12})R_{\bm{q}_{1\bot}}(\theta_{12})
    R_{\bm{q}_1}(-\phi_{12})   R_{z}(\phi_{1})R_{y}(\theta_{1})
    R_{z}(-\phi_{1}) \begin{pmatrix}   0 \\ 0 \\ E_p \end{pmatrix} \end{pmatrix}\,,
\end{align}
and 
\begin{align}
   \epsilon_{\pm 1}=\frac{1}{\sqrt{2}}\begin{pmatrix} 0 \\   R_{\bm{q}_1}(\phi_{12})R_{\bm{q}_{1\bot}}(\theta_{12})
    R_{\bm{q}_1}(-\phi_{12})   R_{z}(\phi_{1})R_{y}(\theta_{1})
    R_{z}(-\phi_{1}) \begin{pmatrix}   \mp 1 \\ -i \\ 0 \end{pmatrix} \end{pmatrix}\,.
\end{align}

\end{onecolumngrid}
\end{document}